\documentclass[submit]{smj}

\usepackage{array}
\usepackage{booktabs}
\usepackage{float}
\usepackage[width=.8\textwidth]{caption}
\usepackage{textcomp, gensymb}
\usepackage[normalem]{ulem}

\providecommand{\tightlist}{%
  \setlength{\itemsep}{0pt}\setlength{\parskip}{0pt}}

\usepackage{setspace}
\usepackage{etoolbox}
\BeforeBeginEnvironment{tabular}{\begin{singlespace}}
\AfterEndEnvironment{tabular}{\end{singlespace}}
\Author{Ryan Peterson\Affil{1}
        and Joseph Cavanaugh\Affil{2}
}
\AuthorRunning{Peterson and Cavanaugh}

\Affiliations{

\item Department of Biostatistics and Informatics,
      Colorado School of Public Health,
      University of Colorado - Anschutz Medical Campus,
      CO, USA

\item Department of Biostatistics,
      College of Public Health,
      University of Iowa,
      IA, USA

}   

\CorrAddress{Ryan Peterson,
             Department of Biostatistics and Informatics,
             University of Colorado - Anschutz Medical Campus,
             13001 E 17 Pl,
             Aurora, CO, 80045, USA}
\CorrEmail{ryan.a.peterson@cuanschutz.edu}
\CorrPhone{(+1)\;303\;724\;0853}

\Title{Fast, effective, and coherent time series modeling using the sparsity-ranked lasso}
\TitleRunning{Sparsity-ranked lasso for time series}

\Abstract{The sparsity-ranked lasso (SRL) has been developed for model selection and estimation in the presence of interactions and polynomials. The main tenet of the SRL is that an algorithm should be more skeptical of higher-order polynomials and interactions \emph{a priori} compared to main effects, and hence the inclusion of these more complex terms should require a higher level of evidence. In time series, the same idea of ranked prior skepticism can be applied to characterize the potentially complex seasonal autoregressive (AR) structure of a series during the model fitting process, becoming especially useful in settings with uncertain or multiple modes of seasonality. The SRL can naturally incorporate exogenous variables, with streamlined options for inference and/or feature selection. The fitting process is quick even for large series with a high-dimensional feature set. In this work, we discuss both the formulation of this procedure and the software we have developed for its implementation via the \textbf{fastTS} R package. We explore the performance of our SRL-based approach in a novel application involving the autoregressive modeling of hourly emergency room arrivals at the University of Iowa Hospitals and Clinics. We find that the SRL is considerably faster than its competitors, while generally producing more accurate predictions.}


\begin{document}

\maketitle

\hypertarget{introduction}{%
\section{Introduction}\label{introduction}}

In time series analyses, the comfort of asymptopia, where our series of interest has infinite observations and can be suitably characterized by a small set of endogenous parameters, can become a fool's paradise. With freely available statistical and computational tools built for high dimensional problems, those who dare to venture away from conventional asymptotic theory may be able to build better predictive models and learn more richly about causal mechanisms at play in a data-saturated world.

In this work, we will show how new applications of existing statistical theories and tools can be utilized in time series analyses to produce better inferences and forecasts, focusing on scenarios with complex seasonal patterns and exogenous covariates. We start by describing the background and reviewing existing methods in this space, including common means of modeling seasonal time series data. We then illustrate how ranked-sparsity-based tools such as the sparsity-ranked lasso (SRL) can be extended to outperform existing methods, proceeding through both the formulation of the procedure and the software we have developed for its implementation via the \textbf{fastTS} R package. We compare SRL-based methods with competitors under several plausible simulation scenarios with various degrees of generating model complexity. In our application, we explore the performance of our SRL-based approach relative to alternative methods on the autoregressive modeling of hourly emergency room visits at the University of Iowa Hospitals and Clinics (UIHC). We conclude with a discussion, including areas of future research.

\hypertarget{background}{%
\section{Background}\label{background}}

In this section, we will describe common models for seasonal time series data, several candidate fitting procedures, and several considerations for choosing parameters for these models.

\hypertarget{autoregressive-models}{%
\subsection{Autoregressive Models}\label{autoregressive-models}}

Consistent and continual data collection is beginning to pervade our lives in unprecedented and unexpected ways. Time series data, where one variable is measured many times at regular intervals, is ubiquitous. A measurement in the present usually depends on its past (say, \(p\)) values to some extent, and this dependence can be well captured by the autoregressive model. An autoregressive model of order \(p\), or an AR(\(p\)) model, can be written in a form analogous to a traditional regression model and estimated using standard ordinary least squares (OLS) techniques. The model has the structure
\[
y_t = \beta_0 + \beta_1 y_{t-1} + ... + \beta_p y_{t-p} + \varepsilon_t
\]
\noindent
where after controlling for the autocorrelation in the conditional mean structure, the residuals are assumed to be independent and normally distributed about zero.

One common question is the best way to select \(p\) in the AR(\(p\)) model in a given application -- many techniques are possible, and some of the more common ones are outlined in the next section. Another related question is how to choose \(p_{\max}\), that is, the highest order \(p\) that is considered in this model selection problem. Many times, \(p\) is rather small, and selected on the basis of visual inspection of autocorrelation function (ACF) and partial autocorrelation function (PACF) plots which display sample autocorrelations and partial autocorrelations on lags up to \(p_{\max}\). However, this technique is imperfect, as it injects human error and the potential for overfitting into the model selection process. Further, the need for visual inspection presents a barrier in higher dimensional settings where automated model selection methods are needed.

A time series may also be expected to follow some pattern (or patterns) of seasonality; for example, online searches related to the National Football League will peak during the football season every year. For this reason, it is important that this seasonality be accomodated in the autoregressive model. Let \(\boldsymbol \phi\) refer to the local parameters and \(\boldsymbol \theta\) refer to the seasonal parameters, and say that we know the periodicity of the seasonality is \(m\) time measurements. The AR model can be written as
\[
y_t = \beta_{0} + \sum_{j=1}^p \phi_j y_{t-j} + \sum_{j=1}^P \theta_j y_{t-jm} + \varepsilon_t
\]
\noindent
where again, the \(\varepsilon_t\) are assumed to be independent and normally distributed conditional on the mean structure. As an illustrative example, if we consider a setting where the seasonal period is yearly, and the series is monthly, then \(m = 12\). If the series is weekly, then \(m = 52\), and so on. This AR model is referred to as a seasonal autoregressive (SAR) model, and has parameter collections of size \(p\) and \(P\), which denote how many local and seasonal components are to be estimated, respectively. As with the standard AR model, model selection must take place in order to select \(P \in \{0, 1, 2, ..., P_{\max} \}\).

The benefit of representing the AR and SAR models in this linear model form is that it becomes clear how least-squares (or some other estimation technique) can be used to estimate the AR parameters. However, there are also models that cannot be easily represented by this lagged model form -- for instance, those with moving average (MA) components. The now-ubiquitous autoregressive integrated moving average (ARIMA) model can handle moving average terms as well as differencing, which may be advisable if the time series is not stationary \citep{jonathan2008time}.

\hypertarget{existing-methods-for-order-selection}{%
\subsection{Existing Methods for Order Selection}\label{existing-methods-for-order-selection}}

For ARIMA models, it is common practice to use a likelihood-based estimation procedure to fit models with potentially both AR and MA terms. After a set of candidate models has been fit, an information criterion, such as Akaike's Information Criterion (AIC), its small sample corrected version (AICc), or the Bayesian Information Criterion (BIC), can be used to select an optimal model order. This process is implemented automatically in the \texttt{forecast} package in R \citep{forecast_pap, forecast_pac}. In this framework, the candidate models can be fit either in a step-wise fashion or in a best-subsets fashion. In either case, we refer to this method as automated-ARIMA.

If only AR and SAR models are considered, then the Least Absolute Shrinkage and Selection Operator (the lasso) \citep{tibs1996} can be used to estimate the coefficients and select an optimal \(p\) and \(P\) simultaneously. An important benefit to this approach is that the autoregressive components can be selected and estimated simultaneously alongside exogenous covariates. In the typical ARIMA selection framework, the need to select from a set of exogenous covariates (denoted by \(X\)) can complicate the model selection process considerably; neither step-wise selection nor best-subsets are feasible when the dimension of \(X\) is large, when \(p_{\max}\) is large, or when the sample size \(n\) is massive.

\hypertarget{complications-in-modeling-periodic-patterns}{%
\subsubsection{Complications in modeling periodic patterns}\label{complications-in-modeling-periodic-patterns}}

In the modeling of time series that exhibit prominent recurring patterns, it can be important to distinguish between cyclic and seasonal periodicity. \emph{Cycles} refer to patterns arising from variable periodicity whereas \emph{seasonality} refers to patterns arising from fixed periodicity. (See, for instance, \citet{hyndman2018forecasting}). Often, cyclic time series can be effectively modeled using traditional, lower-lag autoregressive terms. With seasonal time series, the length of the fixed period \(m\) is generally implied by the context of the underlying phenomenon. For example, \citet{ss2000} describe an environmental series called the Southern Oscillation Index (SOI) that is measured monthly for 453 months ranging over the years 1950-1987. This series exhibits a strong yearly period \((m=12)\), which is expected. However, it also exhibits a cycle corresponding to approximately 50 months, which is possibly due to the El Nino temperature cycles. The authors claim that the monthly SOI series can be explained ``as a combination of two kinds of periodicities, a seasonal periodic component of 12 months and an El Nino component of about three to five years.'\,' For this series, the yearly period would be viewed as seasonal whereas the El Nino component could be viewed as cyclical.

For non-cyclical seasonal time series, especially when the data are collected at very short intervals relative to the length of the series, multiple modes of seasonality may exist. If the series is modeled as an autoregression with a large value of \(p_{\max}\), determining which lags should be represented in the autoregression may translate to a complex variable selection problem. For instance, suppose an analyst is attempting to model an hourly time series collected over multiple years, where substantive periodic patterns may be daily \((m=24)\), weekly \((m = 7 \times 24 = 168)\), or monthly (with \(m\) in the neighborhood of \(30.4 \times 24 = 730\)). Along with each seasonal period \(m\), the accompanying seasonal order \(P\) must be considered. Thus, in determining which lags will be needed to adequately model the series, one must not only consider the values of \(m\) that represent potentially relevant periods, but also various multiples and combinations of these periods (e.g., 24 and 48; 24, 48, and 72; 24 and 168; 24 and 168 and 730; etc.). Additionally, including lags near the pertinent seasonal lags could also be useful; for example, if there are strong lag\(-1\) effects, and strong lag\(-m\) effects, then it is quite likely that there are also lag \(m-1\) or lag \(m+1\) effects. Finally, when one conditions on a seasonal series in the model mean structure (e.g.~temperature or day of week), this may sufficiently capture a seasonal pattern such that seasonal AR terms become unnecessary and may degrade the model performance. Thus, determining which lags should be included in the autoregressive model becomes a non-trivial variable selection problem, often in a high-dimensional context (if \(p_{\max}\) is large).

A separate issue in modeling non-cyclical seasonal time series may occur when the relationship between the seasonal period and the sampling rate is not well-understood. For example, the sound energy from a recorded song likely exhibits seasonality owing to the song's natural tempo, however unless song's tempo is documented, the seasonal period cannot be connected to the sampling rate of the recording. In such cases, it can be difficult to specify a definitive value for \(m\), and alternative data-based methods are required.

\hypertarget{potential-solutions}{%
\subsubsection{Potential solutions}\label{potential-solutions}}

In modeling seasonality in the AR framework, a valuable technique involves the inclusion of nearby seasonal lags as additional parameters:
\begin{align}
y_t = \beta_0 + \sum_{j=1}^p \phi_j y_{t-j}
+ \sum_{j=1}^P \boldsymbol \theta_j^T
  \boldsymbol y_{(t-jm - c):(t-jm + c)} + \varepsilon_t \label{eq:3_1}
\end{align}
\noindent
Here, we employ the notation \(\boldsymbol y_{a:b}^T = (y_a, y_{a+1}, ..., y_{b-1}, y_{b})\) to represent the full set of observations between time \(a\) and time \(b\), and \(\boldsymbol \theta_j\) refers to the vector of length \(2c+1\) of coefficients on lags near \(t-jm\). In words, since it may be unknown which lags near \(t-jm\) are important, this approach includes all the lags at or near \(t-jm\) within \(c\) that may be important in modeling \(y_t\). Note that an AR(\(p\)) model is achieved when \(\boldsymbol \theta_j=\boldsymbol 0 \ \forall \ j\), and a SAR(\(P\)) model is achieved if \(c = 0\).

If we use the lasso to fit this model, it will naturally estimate some coefficients to be zero, so we can re-frame this selection problem, setting \(p_{\max} = m P_{\max} + c \overset {\text {def}} = p^*\). When this simplifying assumption is incorporated into the framework, the seasonal AR model may be viewed as one of many candidate lasso models, and equation \eqref{eq:3_1} simplifies to the potentially high-dimensional local AR model:
\begin{align}
y_t = \beta_0 + \beta_1 y_{t-1} + ... + \beta_{p^*} y_{t-p^*} + \varepsilon_t \label{eq:3_2}
\end{align}
\noindent
In words, instead of assuming there will be structural gaps in the important lags in the SAR model, we parametrize every coefficient up through the maximum possible lag and will take care (in the following section) to ensure the solution is sparse and resembles the SAR model. This formulation means that a very high order AR model could be selected, depending on the expected pattern (or patterns) of seasonality. However, if the primary goal is prediction in large-sample settings, there is no downside to selecting a high order model (other than potentially adding in a lot of noise, which hopefully can be avoided with the use of the lasso and an effective model selection criterion).

Another popular method for fitting seasonal time series with multiple modes of seasonality is called the TBATS method (which stands for ``trigonometric, Box-Cox transform, ARMA errors, trend, and seasonal components'') \citep{de2011forecasting}. As with automated-ARIMA, the TBATS method can be implemented efficiently in the \texttt{forecast} package, but importantly, this software does not allow for TBATS to be used in conjunction with exogenous variables. Further, the mode(s) of seasonality must be pre-specified in this framework, and series with multiple plausible modes may benefit from further model selection.

\hypertarget{methods}{%
\section{Methods}\label{methods}}

In this section, we show how the SRL can be used in the time series regression framework to fit AR and SAR models quickly, effectively, and accurately, while also optionally accommodating and selecting from a set of important exogenous features.

\hypertarget{dynamic-penalty-tuning-with-the-sparsity-ranked-lasso}{%
\subsection{Dynamic Penalty Tuning with the Sparsity-Ranked Lasso}\label{dynamic-penalty-tuning-with-the-sparsity-ranked-lasso}}

When prior informational asymmetry is present among candidate predictors, for instance when selecting from all pairwise interactions and polynomials, \citet{srlpaper} motivated and explored the use of the sparsity-ranked lasso. Its main tenet is that an algorithm should be more skeptical of higher-order polynomials and interactions \emph{a priori} compared to main effects, and hence the inclusion of these more complex terms should require a higher level of evidence. Otherwise, if interactions and polynomials are treated with ``covariate equipoise'', algorithms will tend to yield many more false discoveries for interactions and polynomials, which makes models less transparent, harder to communicate, and worse at prediction. The intuition behind the lasso-based implementation of ranked sparsity (the SRL) involves assigning weights to each candidate covariate based on the degree of skepticism that the covariate should in fact enter the model, as illustrated in the expression below:
\[
\left|\left|\boldsymbol y - X\boldsymbol \beta\right|\right|^2 + \lambda \sum_{j=1}^k w_j | \beta_j|
\]
\noindent Here, \(k\) refers to the column dimension of the matrix of covariates \(X\). The addition of the \(w_j\) is the only difference between this formulation and the traditional lasso \citep{tibs1996}, and while the adaptive lasso (which shares the same representation) fixes the \(w_j\) values in a way that corresponds to a data-based ``first glance'' \citep{adaptivels}, the SRL sets these weights based on a prior degree of skepticism related to each covariate. While the intuition behind these weights appears to be quite subjective for the SRL, in its motivation, the authors justified several objective ways of characterizing this skepticism in the context of interactions and polynomials. This concept can be extended to selecting features in a time series regression framework, which we describe in this section.

The SRL is designed to be of use when the assumption of covariate equipoise, defined as the prior belief that all covariates are equally worthy of entering into the model, is not satisfied. In the AR framework, this assumption is definitely not satisfied, and the SRL can address the resulting challenge. Particularly, going from equation \eqref{eq:3_1} to equation \eqref{eq:3_2} in the previous section is suspect; we have very good reason to believe that the effects on the lags between \(p\) and \(m-c\) are equal to zero. We are also much more inclined to think that more recent lags are more likely to be important \emph{a priori} than lags representing the more distant past. For seasonal series, we are much more willing to believe the \(m^{th}\) lag to be predictive than other periodic lags. The SRL can accommodate such expected differences in skepticism by scaling the penalty differently for different lags in the model. In this subsection, we discuss two methods for adapting the SRL to handle time series data: we can either parametrize skepticism, or we can use the data to inform it.

\noindent
\emph{Parametrizing skepticism}

\begin{figure}

{\centering \includegraphics[height=5.1in]{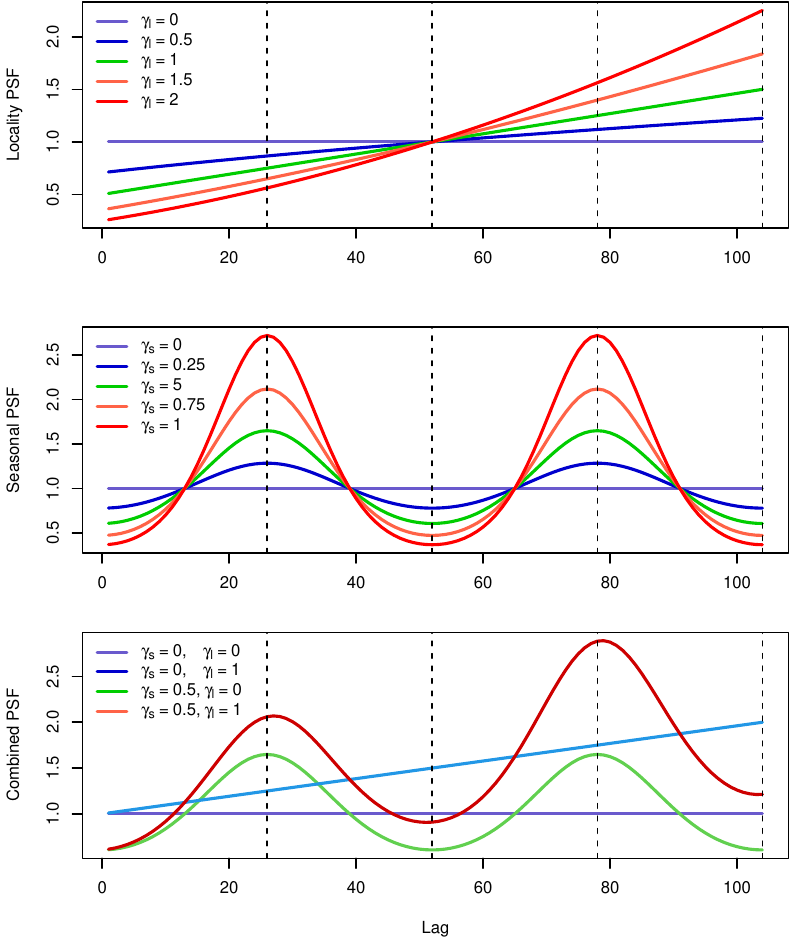} 

}

\caption{Parametrized penalty scaling functions (PSF) for coefficients on lags up to $p^*$ in a time series SRL setting. The top plot refers to the locality PSF ($c=.5$), the middle plot refers to the seasonal PSF, and the bottom plot is their combined PSF. In the bottom plot, $c=1$.}\label{fig:fig01}
\end{figure}

Since we have good reason to believe that recent and seasonal lags are more likely to be important, but we are not certain, we can operationalize this notion using what we call \emph{penalty-scaling functions}. These penalty-scaling functions, which we denote as \(f\), provide ways of informing the weight of the lasso penalty \(w_j\) that corresponds to the \(j^{th}\) lag, i.e.~\(w_j = f(j, \cdot)\). We discuss two different penalty-scaling functions: one that relates to a lag's ``locality'' (i.e.~its recency), and one that relates to its seasonality. We also show how combinations of these two penalty-scaling functions can be utilized in a manner that is both flexible and widely applicable. For locality, we use \(\gamma_l\) to illustrate the strength of this scaling factor, which is treated as a tuning parameter. For \(\gamma_l \geq 0\), and for a coefficient on the \(j^{th}\) lag of \(y_t\), we define the local penalty-scaling function as
\[
f_{l}(j, \gamma_l) = (j / p^* + c)^{\gamma_l} = \exp\left\{\gamma_l \log (j/p^*+c)\right\}
\]
\noindent
If \(\gamma_l = 0\), the resulting procedure is equivalent to the ordinary lasso, since the function evaluates to 1 \(\forall \ j\). The addition and the choice of the constant \(c\) is somewhat arbitrary. If \(c=.5\), this ensures that the scaling factor evaluates to 1 at the middle lag considered, \(p^*/2\). This choice of \(c\) effectively relaxes the original lasso's penalty for early lags, and increases it for later lags. One could also set \(c=1\) instead, which would simply increase the penalty for all lags instead of having the relaxing property. In practice, since one has to tune the original lasso penalty \(\lambda\) anyway, the choice of \(c\) typically makes little to no difference. This function is visualized with \(c=.5\) for various \(\gamma_l\) in the top plot of Figure\(~\)\ref{fig:fig01} for all lags up to \(p^*\).

For the seasonal penalty-scaling function, we use the parameter \(\gamma_s\) to represent the strength of the penalty scaling function. For \(\gamma_s \geq 0\), the proposed penalty scaling function for the \(j^{th}\) lag is
\[
f_{s} (j, \gamma_s, m) =
\exp\left\{\gamma_s \left[-\cos\left(\frac{2\pi*j}{m}\right)\right] \right\}
\]
\noindent
Here \(m\) is the (suspected) seasonal frequency, which is set to 52 in the visualization in the middle plot of Figure\(~\)\ref{fig:fig01}. Similar to the locality function, setting \(\gamma_s\) to 0 yields the ordinary lasso. Finally, we can combine the two penalty-scaling functions by simply multiplying \(f_l\) and \(f_s\) together. For \(\gamma_l \geq 0, \gamma_s \geq 0\), we have
\[
\begin{aligned}
f_{ls} (j, \gamma_s, \gamma_l, m) &=
 \exp\left\{\gamma_s \left[- \cos\left(\frac{2\pi*j}{m}\right)\right]
\right\}
*  (j / p^*+ c)^{\gamma_l}\\
&= \exp\left\{\gamma_s \left[- \cos\left(\frac{2\pi*j}{m}\right)\right]
+ {\gamma_l} \log (j / p^* + c)
\right\}
\end{aligned}
\]
\noindent
This penalization is visualized for different values of \(\gamma_l\), \(\gamma_s\), and \(j\) in the bottom plot of Figure\(~\)\ref{fig:fig01}.

The parametrically-weighted SRL is especially useful for handling gaps in the AR structure, and it can be straightforwardly extended to considering multiple modes of seasonality. However, the method is not without drawbacks. A minor drawback occurs when the seasonal frequency \(m\) is definitive (without multiple modes), and yet the algorithm selects lags close to \(m\) instead of the ``true'' model with only the \(m^{th}\) lag included (e.g.~for simulated SAR(1) series). SRL penalizes the lag-\(m\) coefficient very similarly to those surrounding \(m\), and the lasso is prone to select from correlated variables somewhat arbitrarily. However, since prediction is usually the aim, and real data arguably do not have ``true'' lags, this drawback is quite minor as either provides a useful approximation. Finally, the combined penalty scaling function is not always defensible; the period of the wave depends (to a small degree) on the locality hyperparameter, and therefore the two \(\gamma\) parameters interfere with each other. This can lead to issues of identifiability; we have found in practice that the optimal solution using this technique, as measured by an information criterion, can lie on a line of \(\gamma_l, \gamma_s\) tuning parameters, although this line often does not intersect with the lasso solution (where both \(\gamma_l = 0\) and \(\gamma_s = 0\)).

We have explored these parametrized penalty-scaling functions and applied them successfully in multiple applications, and will return to them in the simulation section. However, the SRL method that we discuss in the next subsection performs similarly and is more broadly applicable than the parametrized version.

\noindent
\emph{SRL with the partial autocorrelation function}

Instead of having a predefined parametrized ranking of skepticism for various lags in a time series model, we could use the data to inform these penalty weights. A useful means to accomplish this is to employ the partial autocorrelation function (PACF), defined as the correlation between \(y_t\) and \(y_{t-k}\) after removing the effect of the intervening variables \(y_{t-1}, y_{t-2}, ..., y_{t-k+1}\). The PACF is easy to calculate, and it provides a data-driven measure of the importance of each lag -- many statistical practitioners inspect the PACF to determine which lags should be considered and estimated in an autoregressive model \citep{jonathan2008time}.

The adaptive lasso \citep{adaptivels} involves penalty weights \(w_j\) that are informed by an initial ``first-stage'' estimate of \(\boldsymbol \beta\), often accomplished via a simple OLS estimate. In our case, we suggest using PACF estimates instead by setting
\[
w_j = \left(\frac{1}{|\hat \phi_j|}\right)^\gamma
\]
\noindent
Here \(\hat \phi_j\) represents the estimated PACF on the \(j^{th}\) lag of \(y_t\). Each of these estimates is equivalent to the \(j^{th}\) AR coefficient estimate in an AR(\(j\)) model, and can therefore be obtained via a heuristically-guided solution path from an AR(1) to an AR(\(p^*\)). In words, \(\phi_j\) measures the relative importance of the \(j^{th}\) lag, conditional on all of the more recent lags. Penalty weights estimated with such an approach will tend to be smaller for more recent lags compared to those based on the estimated AR coefficients of the full AR(\(p^*\)) model. Penalty weights estimated using the AR(\(p^*\)) model coefficients do not perform as well (results not shown). We refer to this approach as the SRLPAC (sparsity-ranked lasso with partial autocorrelation) procedure, pronounced ``SRL-pack''.

If there are exogenous covariates, we can include weights for these as well; let \(\boldsymbol w_{endo}\) refer to the vector of weights on endogenous features (i.e., the lags of \(\boldsymbol y\)), and \(\boldsymbol w_{exo}\) refer to that for a set of exogenous features in a \(n \text{ x } k\) model matrix \(X\). The latter exogenous weights can be applied set to the marginal OLS estimates or the OLS estimates controlling for the autocorrelation in the series in a similar adaptive lasso fashion. This approach will share similar properties to the adaptive lasso, and will therefore be useful and computationally feasible even with high dimensional (high \(k\)) feature sets. However, estimation shrinkage complicates formal statistical inference on coefficients. Alternatively, if \(\boldsymbol w_{exo}\) is set to \(\boldsymbol 0\), then the coefficients on exogenous features are unpenalized, and formal means of statistical inference on them is possible via ordinary least squares using the penalized linear predictors from the endogenous features as an offset. Regardless of the weights chosen, we refer to this approach as the SRLPACx procedure, pronounced ``SRL-packs,'' to mirror the SARIMAx and ARIMAx approaches which also allow for exogenous variables to be specified. Similarly, we will refer to the parametric version of the SRL for time series described in the previous section as SRLPAR, and SRLPARx when exogenous covariates are included.

The SRLPAC approach is ideal for modeling time series data with complicated seasonality; it is quick, intuitive, and can be conveniently tuned using AICc or BIC provided the sample size is large, which is typically the case in time series settings. Further, SRLPAC does not require the pre-specification of seasonal modes at all; it does this naturally and automatically using the PACF. Finally, the SRLPACx approach has all the same advantages while also allowing for seamless integration of (possibly very many) exogenous variables.

\hypertarget{tuning-parameters}{%
\subsection{Tuning Parameters}\label{tuning-parameters}}

Generally, the hyperparameter for the lasso (and the SRL) models can be tuned using either information criteria (AIC, BIC, or AICc) or cross-validation (CV). BIC will generally select sparser (lower order) autoregressive models, while AIC, AICc, and CV will attempt to minimize prediction error (typically by allowing more coefficients into the model). Since predictive accuracy is usually of primary importance, we recommend using CV or AICc in most cases. With time series data, CV can be accomplished with the expanding window CV, the sliding window CV, or the \(k\)-fold CV procedure outlined in \citet{bergmeir2018note}. As the SRLPAC procedure involves computation of partial autocorrelations which are used for the weights in the subsequent lasso steps, a CV procedure must nest the computation of the PACFs; otherwise we would expect CV error estimates to be biased/optimistic. A similar issue has arisen in the context of the adaptive lasso (e.g. \citet{alassoCV}).

\hypertarget{measuring-predictive-accuracy}{%
\subsection{Measuring Predictive Accuracy}\label{measuring-predictive-accuracy}}

In order to measure prediction accuracy in a time series setting, there are many possible options. Since our methods use least squares to fit parameters, we use the root-mean-squared prediction error (RMSPE) and \(R^2\), estimated using out-of-sample data, as defined in this section. For a set of predicted values \(\hat {\boldsymbol y}^*\), and a set of \(n^*\) out-of-sample observations that were not used in calculating these predictions, \(\boldsymbol y^*\), we define RMSPE as \[\text{RMSPE} = \sqrt{\frac{1}{n^*} \sum_{i=1}^{n^*} (\hat y_i^* - y_i^*)^2 }.\]
\noindent
RMSPE is interpretable in the same unit of measurement as the original measurement for \(y\), and it signifies how far away from the true value our predictions were, on average. We also measure \(R^2\), defined as \[R^2 = 1 -\frac{\sum_{i=1}^{n^*} (\hat y_i^* - y_i^*)^2}{\sum_{i=1}^{n^*}  (y_i^* - \bar y^*)^2}.\]
\noindent
This definition of \(R^2\) differs slightly from others in that there is no guarantee that it be non-negative, as is the case for \(R^2\) in its typical in-sample setting. Still, values close to 1 indicate very accurate predictions relative to naively guessing the mean, and values close to 0 indicate that using the mean of the responses for forecasting would have done as well as the model-based predicted values. Negative values indicate that the model-based predictions actually performed worse than the mean.

\hypertarget{the-fastTS-package}{%
\subsection{\texorpdfstring{The \textbf{fastTS} Package}{The fastTS Package}}\label{the-fastTS-package}}

We have developed the \textbf{fastTS} package in the R statistical software to implement the novel methods described previously. Using the \texttt{fastTS} function, a user must only supply the time series of interest as the outcome, and can optionally supply additional components which include: a matrix of exogenous features, the maximum lag considered (\(p_{\max}\)), a vector of coefficient weights for the endogenous features (\texttt{w\_endo}), a vector of coefficient weights for the exogenous features (\texttt{w\_exo}), a vector of candidate exponents for the penalty weights (\(\gamma\)), a proportion to use for the training of the model if out-of-sample prediction accuracy is of interest, and several other minor components (see package documentation). The ``coefficient weights'' here refer to the inverse of the penalty factor weights in the SRL fitting procedure such that a coefficient with a weight of 0 will have zero chance of entering the model, and a weight of \(\infty\) will be unpenalized. An appealing property that results from setting coefficient weights to \(\infty\), especially for exogenous features, is that statistical inference can be conducted using traditional methods without correcting for post-selection inference or accounting for bias due to the shrinkage of the coefficients. For this reason, the default option for \(\boldsymbol w_{exo}\) is ``unpenalized'', so that coefficients and standard errors are easily estimated and unbiased (via the \texttt{summary} function). The fitting engine for the lasso is accomplished via the \textbf{ncvreg} package \citep{breheny2011} which uses coordinate descent, and supports additional features such as nonconvex penalization, an L2 penalty parameter which allows for the elastic net, and the modeling of non-normally distributed outcomes. Model fitting involves two tuning parameters, \(\lambda\) (the extent of overall coefficient penalization) and \(\gamma\) (the degree of emphasis placed on the weights), as defined previously. By default the \texttt{fastTS} function will use the AICc to select the best values, starting with \(\gamma \in \left\{0, 0.25, 0.5, 1, 2, 4, 16 \right\}\) and running each of these through 101 decreasing values of \(\lambda\) according to the standard limits implemented in \textbf{ncvreg}. BIC is also available as an alternative means of tuning parameter selection, although the package stops short from implementation of any CV procedure discussed in Section 3.2 (these would be welcome future contributions).

The \textbf{fastTS} package is currently available on \href{https://github.com/petersonR/fastTS}{GitHub} and on the Comprehensive R Archive Network. A tutorial on several popular time series data sets is included in the package website as a vignette, as well as a more in-depth vignette which more completely describes the programmatic details of the SRLPAC detailed in the forthcoming application.

\hypertarget{simulation-study}{%
\section{Simulation Study}\label{simulation-study}}

\hypertarget{simulation-set-up}{%
\subsection{Simulation Set-up}\label{simulation-set-up}}

To demonstrate the performance of our proposed method and its competitors, we have implemented a set of Monte Carlo simulations under controlled settings. We consider five generative data models based on fitted models for the emergency room visit data set described in our application. Briefly, in our application, we suspect multiple modes of seasonality and potentially important exogenous covariate effects which include concurrent hourly temperature and a set of holiday (and day-after holiday) indicator variables. Our main generative settings were designed to capture various degrees of simplicity: MA(\(q\)=5), SARMA(\(p\)=1, \(q\)=1, \(P\)=1, \(Q\)=1, \(m\)=24), MAx(\(q\)=5), and SARMAx(\(p\)=1, \(q\)=1, \(P\)=1, \(Q\)=1, \(m\)=24). Additionally, we include a data generating model based on an AR fit with ten local lags and one seasonal lag, where local AR coefficients (lags 5-7), which were estimated to be less than 0.05 in absolute value, were set exactly to zero. This latter setting, which we refer to as the SAR gap model (SARx gap model with covariates), simulates what occurs when gaps exist in the local AR structure.

Due to the computational time needed to fit automated ARIMA and TBATS methods, we reduced the sample size considerably to \(n=500\) time points (from \(n=41,640\) in the application). For the generation of the covariates, we sampled \(n=500\) rows without replacement from the full original covariate design matrix. If any of the sampled covariates had zero variance, we removed them from the model matrix prior to fitting models; this could occur, for example, if there were no Thanksgiving holidays observed in the sample. We also increased the signal from exogenous terms by multiplying coefficients by a factor of 5 to make their effects easier to detect in a smaller sample. The resulting simulation parameters are listed in Tables\(~\)\ref{tab:tab01} and S1.

\begin{table}

\caption{\label{tab:tab01}Data-based generating model parameters for MA-based simulations.}
\centering
\begin{tabular}[t]{lrrrr}
\toprule
Term & Model 1 & Model 2 & Model 3 & Model 4\\
\midrule
MA 1 & 0.37 & -0.74 & 0.36 & -0.69\\
MA 2 & 0.34 &  & 0.33 & \\
MA 3 & 0.29 &  & 0.29 & \\
MA 4 & 0.22 &  & 0.21 & \\
MA 5 & 0.13 &  & 0.13 & \\
AR 1 &  & 0.79 &  & 0.73\\
SAR 1 &  & 1.00 &  & 1.00\\
SMA 1 &  & -0.97 &  & -0.97\\
10-degree (F) &  &  & 1.65 & 0.66\\
Christmas &  &  & -5.90 & -8.33\\
Christmas + 1 &  &  & 5.03 & 3.09\\
New Years Eve &  &  & -1.01 & -3.18\\
New Years Day &  &  & 2.65 & -0.92\\
Thanksgiving &  &  & -5.34 & -8.82\\
Thanksgiving+1 &  &  & 0.61 & 0.01\\
Independence Day &  &  & -5.93 & -4.66\\
Hawkeye Gameday &  &  & -1.40 & -0.46\\
\bottomrule
\end{tabular}
\end{table}

To fit simulated data produced from these generating models, we used automated SAR, automated SARx, TBATS, SRLPAR, SRLPAC, SRLPARx, and SRLPACx. We restricted our ARIMA-based methods to AR-only models so that the MA-based true generating models could not be perfectly specified by any of the fitting methods. For the SRLPAR method, we set penalty scaling functions using \((\gamma_l, \gamma_s) = (0.1, 0.01)\) to reflect a minor preference toward more recent lags and a seasonal period of \(m\)=24, the latter of which was also specified for SAR, SARx, and TBATS. Finally, in order to assess the quality of the methods, we produced a test set of the same size, then computed each model's predictions on the new data set to obtain the model's RMSPE. We also fit an ``oracle'' model, and we present the median across \(S=10000\) simulations of the relative increase in the RMSPE for each method relative to this oracle. We utilize the median and interquartile range (IQR) rather than the mean because TBATS produced a small number of badly performing models, which influences estimates of its mean performance considerably. All simulation code is included as an appendix.

\hypertarget{simulation-results}{%
\subsection{Simulation Results}\label{simulation-results}}

\begin{figure}

{\centering \includegraphics[height=4.5in]{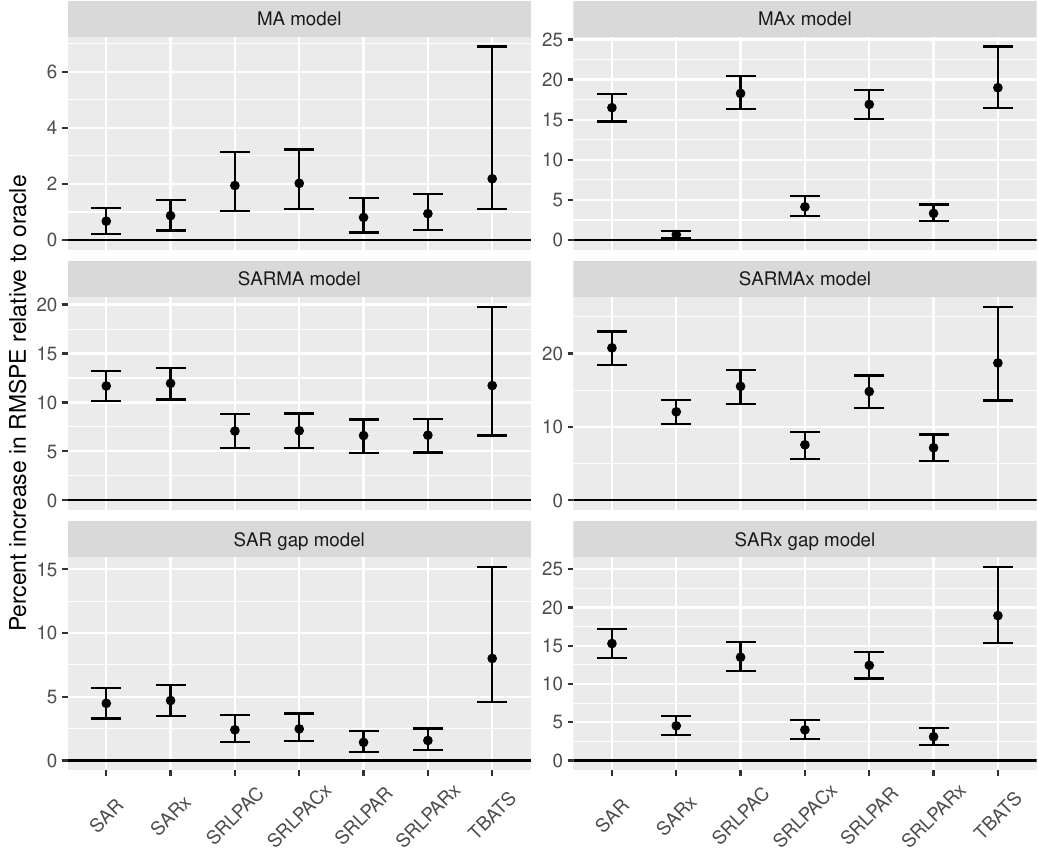} 

}

\caption{Performance of various fitting methods relative to the oracle (true) model for simulations. A value of 1 indicates that the RMSPE is as good as the oracle fit to the same training data. Points represent medians, and bars represent the interquartile range.}\label{fig:fig02}
\end{figure}

\begin{figure}

{\centering \includegraphics[height=4.5in]{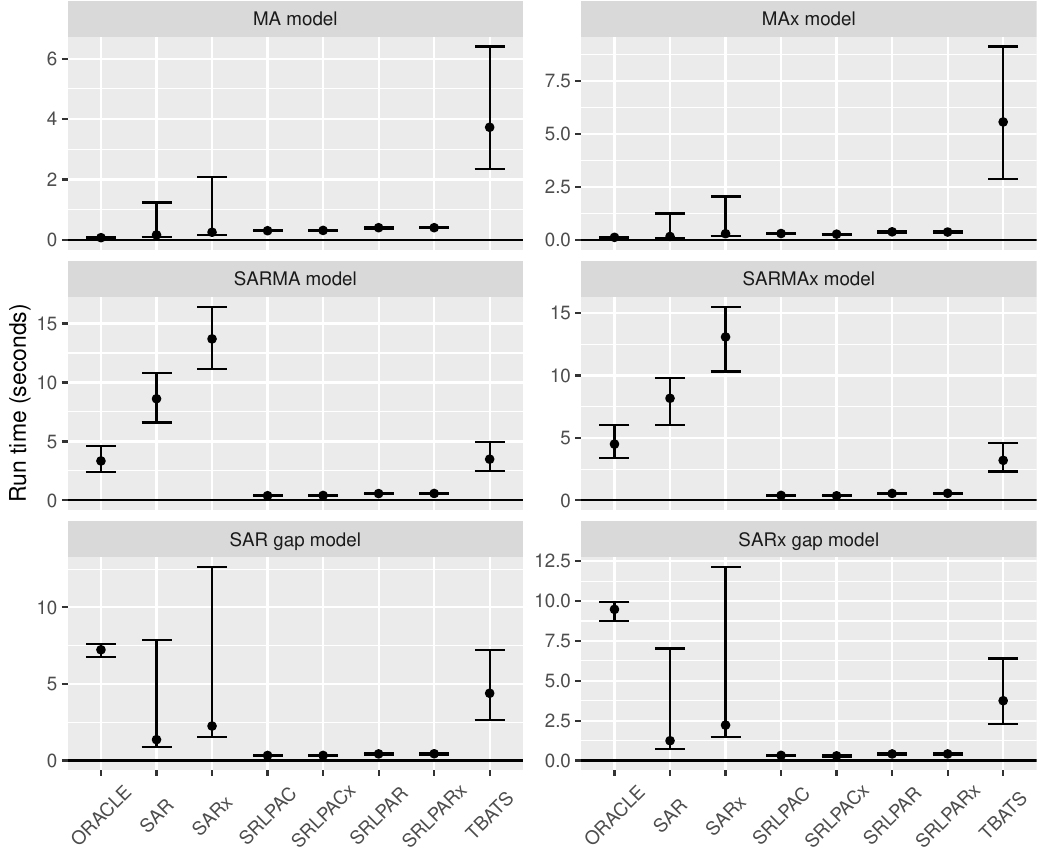} 

}

\caption{Median computation time for fitting methods to 500 simulated data points. Functions used included the \texttt{Arima()}, \texttt{auto.arima()}, and \texttt{tbats()} functions from the \texttt{forecast} package, and \texttt{fastTS()} function from the \texttt{fastTS} package. Points represent medians, and bars represent the interquartile range.}\label{fig:fig03}
\end{figure}

The median and IQR performances and computation times for our simulations are presented in Figures \ref{fig:fig02} and \ref{fig:fig03} respectively.

\captionsetup{width = .98\textwidth}

In the MA(5) data-generating model, SAR- and SRL-based methods perform similarly (with SRLPAC method performing slightly worse than SRLPAR). Models were able to achieve much better predictions when incorporating the exogenous covariates, at which point the difference between SRLPACx and SRLPARx became minor, and both showed a slightly worse performance than the automated SARx model. In the SARMA data-generating model, the TBATS and SAR-based methods performed much worse than those based on SRL. In the SARMAx generating model, the SRLPACx and SRLPARx models performed similarly, better than the automated SARx model, and much better than the other methods which did not account for the exogenous signals. In the SAR gap data-generating models, we see superior performance of the SRL-based methods, with the parametric variant exhibiting slightly better performance than the PACF variant.

The SRL-based methods achieved a very considerable improvement in speed relative to the automated SARx, TBATS, and even the oracle fitting method. The benefits of improved computational speed were more apparent for the more complex data-generating models. For the SAR gap generating model, the automated SAR and SARx fitting methods reflected a high degree of variability in the run times, while the run times for the SRL-based methods were not only fast, but also exhibited little variability.

In short, we have confirmed that the SRL-based methods are well-suited to seasonal data, can fully leverage signal from exogenous features to improve predictions, and are more computationally efficient and consistent than currently available software.

\hypertarget{application-emergency-room-visits}{%
\section{Application -- Emergency Room Visits}\label{application-emergency-room-visits}}

We showcase our methodology in a novel approach to emergency room visit forecasting. For planning purposes, it is highly desirable to accurately forecast the expected number of visits in the emergency room (ER) each hour. Accurate forecasting (and planning) reduces the costs and frustrations associated with having to ``call-in'' extra help. Previous research has indicated that time series models can be developed and utilized to predict ER visits on the daily time-horizon \citep{jones2008forecasting}. However, these daily forecasts, while informative, are less helpful from a practical standpoint since shifts are typically set for under 12 hours. It is useful to have more granular predictions so that personnel resources can be allocated more efficiently on a shift-by-shift basis. Other research investigating the forecasting of ER arrivals at the hourly level does exist, and mostly utilize seasonal ARIMA models to fit the hourly series for a single ER. A detailed literature review is outside the scope of this manuscript.

In this study, we examine hourly visit counts for the Emergency Department at the University of Iowa Hospital and Clinics (UIHC) from July 2013 to March 2018. Since this is hourly data over a long time span, the sample size in this modeling problem is large (\(n = 41,640\)). The high sample size also yields minor computational concerns for some of the methods we have described. Multiple modes of seasonality are feasible; we expect to see more visits during the day than at night, and weekly and yearly cycles are clearly plausible. Monthly seasonality could be exhibited if, for example, ambient light (due to lunar cycles) is associated with ER visits. ER physicians have long believed a ``full moon'' effect is feasible (e.g. \citet{zargar2004full}).

\begin{figure}

{\centering \includegraphics[height=2.75in]{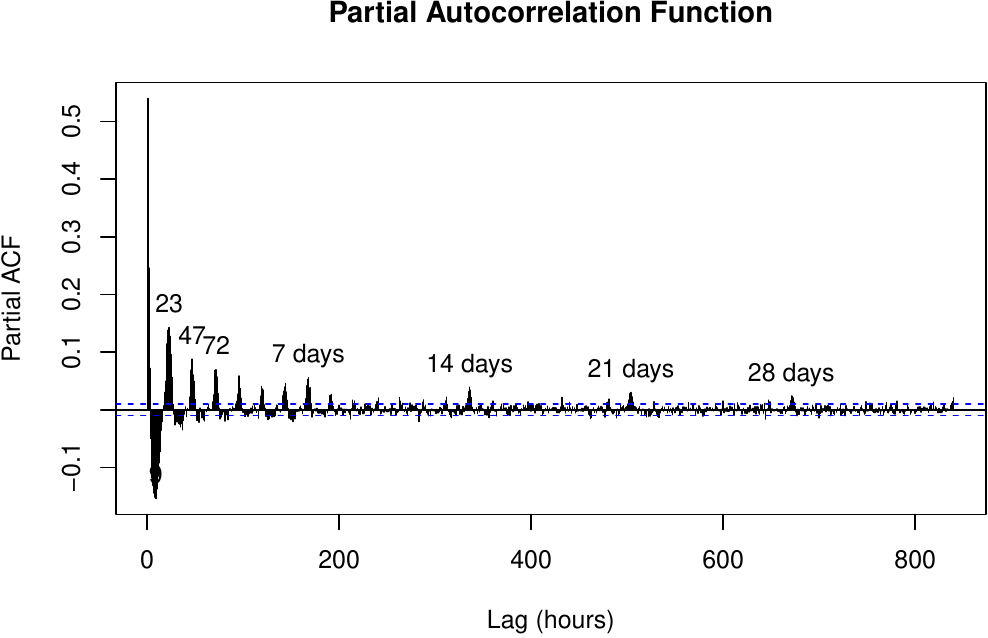} 

}

\caption{Partial Autocorrelation Function for hourly emergency room arrivals.}\label{fig:fig04}
\end{figure}

In Figure\(~\)\ref{fig:fig04}, we indeed observe many modes of seasonality. The most prominent correlation is the AR(1) term, but there are also large spikes around 23 hours, 47 hours, 72 hours, 7 days, 14 days, 21 days, and 28 days.

We consider monthly effects, holiday effects, and concurrent hourly temperature in Iowa City, IA, as exogenous features, unpenalized in the regression. For months and holidays, we used indicator variables to denote month as well as specific holidays, including Christmas, New Year's Eve, New Year's Day, Thanksgiving, Independence Day, and Hawkeye game-day (i.e., the day of an Iowa Hawkeye college football game). We also include holiday ``plus-one'' effects for Christmas and Thanksgiving to examine if there are lingering effects of these major holidays.

Before fitting the models, we divide our sample into a training set (the first 90\% of data, 7/1/2013-10/9/2017) and a test set (the final 10\% of data through 3/31/2018). Using the training set, we fit the following models:

\begin{enumerate}
\def\labelenumi{\arabic{enumi})}
\tightlist
\item
  SRLPAC
\item
  SRLPAC with exogenous variables (SRLPACx)
\item
  Automated SARIMA, with \(p\), \(q\), \(P\), and \(Q\) selected using AICc. As has been noted, only one mode can be supplied, so we specified a daily frequency (\(m=24\)) for this setting based on the PACF plot.
\item
  TBATS model with daily, weekly, monthly, and yearly frequencies specified: \(m = (24, 168, 730, 8766)\).
\end{enumerate}

In addition to computing the metrics outlined in the previous section on the test data, we repeat the fitting process 12 times to calculate the mean computation time (MCT) for each method. For both SRLPAC models, we use AICc to select a value of \(\lambda\) and \(\gamma\). For the SRLPACx model, we interpret terms in context as well as 95\% confidence intervals based on OLS models where the penalized linear predictors based on the endogenous variables are included as an offset. Finally, for staffing purposes, it is arguably more informative to investigate how well the model performs at predicting the number of visits that occur in a given 8-12 hour window, rather than at the hourly level. So, we also investigate the performance of the SRL and TBATS methods in predicting the 10-hour rolling sum of patients, which was calculated using the sum of the 1- through 10-step-ahead predictions on the test data set.

\begin{table}

\caption{\label{tab:tab02}Model performance for hourly arrivals. Prediction metrics were estimated using a left-out test sample, and the mean computation time (MCT) in minutes was computed across 12 replications on a single core of a machine running Mac OS.}
\centering
\begin{tabular}[t]{lrrr}
\toprule
  & $R^2$ & RMSPE & MCT (Minutes)\\
\midrule
SRLPAC & 0.528 & 2.651 & 0.85\\
SRLPACx & 0.530 & 2.645 & 1.09\\
SARIMA & 0.374 & 3.052 & 6.42\\
TBATS & 0.529 & 2.646 & 8.50\\
\bottomrule
\end{tabular}
\end{table}

We see in Table\(~\)\ref{tab:tab02} that the automated SARIMA approach performed the worst by all measures; it took a relatively long time to run, and the models it produced did not predict accurately. The TBATS method did predict one-step ahead quite well, though it also took much longer to run than either SRLPAC approach. SRLPAC and SRLPACx performed well, both with and without the exogenous variables, the inclusion of which seems to offer a modest improvement in predictive accuracy. The SRL with exogenous variables and TBATS were essentially tied and performed best at one-step predictions.

\begin{figure}

{\centering \includegraphics[height=4in]{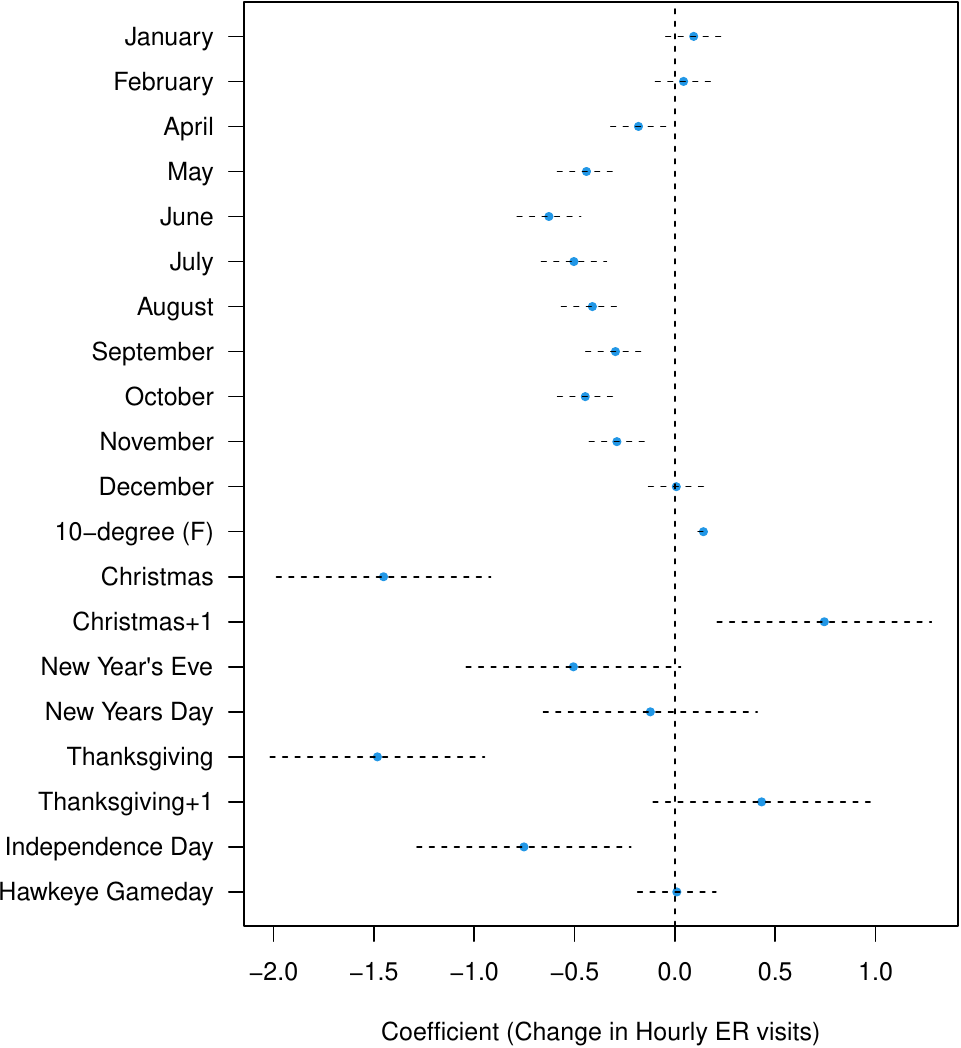} 

}

\caption{Coefficient estimates and 95\% confidence intervals for exogenous variables in SRLPAC model. Values refer to the expected change in hourly visits controlling for other factors in the model, including temporal correlation.}\label{fig:fig05}
\end{figure}

In contrast to TBATS, we can make interpretable inferences regarding the exogenous variables using the SRL-PACx model, where interpretations condition on the rest of the covariates in the model as well as \emph{relevant}\footnote{Relevant local and periodic autocorrelation in the series is captured by SRLPAC approach. The selected estimates are visualized in Figure$~$S2.} autocorrelation (Figure\(~\)\ref{fig:fig05} and Table\(~\)S3). We find that hourly ER visits exhibit seasonality on a yearly-scale in addition to a very strong positive association concurrent hourly mean temperature; for every 10 degree F increase in temperature, we expect to see about 0.14 more patients per hour. We observe very strong negative effects on the major holidays, most of all Christmas and Thanksgiving on which there were about 1.5 fewer patients per hour, on average. This effect seems to have consequences; on the day after Christmas and Thanksgiving, there are about 0.74 and 0.43 more patients per hour, respectively. The effects of New Year's Eve, New Year's Day, and Independence Day are all negative, but are less pronounced than Christmas and Thanksgiving. There does not seem to be a large change in the number of patients per hour on Hawkeye game-days.

\captionsetup{width = .9\textwidth}

Table\(~\)\ref{tab:tab04} shows model prediction performance for how many patients arrived in the ER in a 10-hour window. The SRLPAC methods performed similarly to one another with a slight advantage toward the SRLPACx model, with both methods outperforming TBATS and automated SARIMA. This optimal model was able to describe 85.4\% of the variation in the 10-hour rolling sum of ED arrivals. The SRLPACx model's RMSPE was the lowest; its prediction was on average only about 8.6 patients off of the true value. Figure\(~\)S1 shows the accuracy of the SRLPACx model in predicting the 10-hour rolling counts. Of note, the SRLPAC and SRLPACx models generated 1- through 10-step ahead predictions in under 5 minutes, while TBATS and SARIMA procedures took considerably longer (2 hours for the TBATS predictions and over 6 days for the SARIMA predictions).

\captionsetup{width = .525\textwidth}

\begin{table}

\caption{\label{tab:tab04}Model prediction performance for 10-hour rolling sum of patient arrivals.}
\centering
\begin{tabular}[t]{lrr}
\toprule
  & $R^2$ & RMSPE\\
\midrule
SRLPAC & 0.839 & 9.004\\
SRLPACx & 0.854 & 8.564\\
SARIMA & 0.761 & 10.995\\
TBATS & 0.826 & 9.375\\
\bottomrule
\end{tabular}
\end{table}

\hypertarget{discussion}{%
\section{Discussion}\label{discussion}}

When time series data exhibit complex and potentially multiple-mode seasonality as well as exogenous covariates, popular existing modeling strategies lack crucial flexibility. We have shown that the sparsity-ranked lasso is adaptable to time series data in several ways. Not only do SRL-based methods rival the predictive accuracy of other state-of-the-art forecasting methods, they facilitate the seamless incorporation of exogenous data and run considerably faster for large data sets. Joining the sparsity-ranked lasso with the partial autocorrelation function (SRLPAC), we were able to accurately forecast the number of patients who would arrive at the UIHC Emergency Department to within an average of about 8.5 patients per 10-hour time window. We were also able to use the SRLPACx model to make inferences about patterns in the data, finding large associations between ER arrivals and temperature, month, and holidays.

For similar staffing purposes (and more generally), a quantile of the forecast distribution may be preferred to the mean. For this goal, we note that forecasts should theoretically be normally distributed asymptotically, so the analyst could compute prediction intervals using the estimated forecast variance based on its parametric form for an AR model.

Other researchers have shown that SARIMA models are effective in forecasting the number of ER arrivals for daily data, see \citep{jones2008forecasting} for example. However, we have shown that for local hourly data, the SARIMA model does not perform effectively compared to other methods, and the only methods that can predict well and handle exogenous covariates are SRL-based.

We are not the first to apply the lasso in time series applications (e.g. \citet{Chernozhukov}), though we are unaware of existing methods or software devoted specifically to applying it to seasonal time series. Existing variable selection methods for time series models using information criteria typically make use of recency, dropping older lags first as a part of the tuning process. Relative to this practice, our methods have an advantage when the lag structure includes some kind of ``delayed effect'' (i.e.~there are gaps in the lag structure). We have shown how SRL-based methods can capture this, whereas traditional recency-guided methods using information criteria will not. We expect the SRL-based methods to also well-capture gaps in seasonal lag structures; e.g.~when the first and third seasonal lag are important, but the second one is not.

Although we have only shown empirical evidence in this work, we suspect that the SRL-based methods will perform similarly well for other time series data sets with complex seasonality and/or exogenous features. To this end, we have created the \textbf{fastTS} R package for open-source dissemination, and we hope that other time series modelers find our method to be both useful and accurate. Especially in complex generating settings, our simulations showed our methods to be consistently faster computationally and better at forecasting compared to competing methods. We also found that SRLPAC tends to perform similarly to SRLPAR in complex settings, but SRLPAR performed better when the model was less complex, in which settings SRLPAC's use of the PACF in lieu of specifying \(m\) led to slightly worse performance. In future work, we will compare the SRL-based approaches to other popular forecasting methods in this domain, such as neural networks. We will also investigate model averaging approaches which blend multiple values of the tuning parameters of SRLPAC (\(\gamma\) and \(\lambda\)) based on their likelihoods to provide a potentially more accurate prediction.

\hypertarget{acknowledgements}{%
\section{Acknowledgements}\label{acknowledgements}}

We'd like to thank two anonymous reviewers as well as Dr.~Gary Grunwald for providing vital advice on how to describe the utility of this method in various types of complex seasonal data.

\bibliography{bibliography.bib}

\end{document}